\documentclass[frenchb,oneside,12pt]{article}
{\topskip 2cm}
\usepackage{babel,indentfirst}
\usepackage{latexsym}
\usepackage{amsfonts}
\usepackage[dvips]{graphicx}
\begin{document}
\title{Geometrical and Algebraic structures in Quantum Hall
Systems}
\author{\large J.C. Wallet{\footnote{e-mail address: wallet@ipno.in2p3.fr}}}
\maketitle
\begin{center} 
{Groupe de Physique Th\'eorique, Institut de Physique Nucl\'eaire,
F-91406, ORSAY CEDEX (France)}
\end{center}
\vskip 1 true cm
{\bf{Abstract}}{\footnote{Talk given at the "Conference on Higher Dimensional
Quantum Hall Effect, Chern-Simons Theory and Non-commutative Geometry in
Condensed Matter Physics and Field Theory", Trieste, 1-4 March 2005.}}
: We review the main features of a mathematical framework encompassing some
of the salient quantum mechanical and geometrical aspects of Hall systems
with finite size and general boundary conditions. Geometrical as well as
algebraic structures controlling possibly the integral or fractional
quantization of the Hall conductivity are discussed.
\par
\vskip 4 true cm
PACS: 71.10.-w; 02.40.-k

\strut\thispagestyle{empty}
\pagebreak
\setcounter{page}{1}

%tableofcontents
%listoffigures
%\listoftables
\section{Introduction}
The integral quantization of the Hall conductivity $\sigma_H$ can be
understood as stemming from topological features underlying the Quantum
Hall systems {\cite{PRANG}}. The pioneering argument in that direction
proposed by Laughlin \cite{LAUGH} for a system with a cylindrical geometry
has been refined in subsequent works dealing with non interacting electrons
in a periodic potential \cite{THOUL1} and further extended to take into
account disorder and/or electron mutual interaction \cite{THOUL2},
\cite{WU}, \cite{AVRON1}, \cite{THOUL3}. Soon after \cite{LAUGH}, it has
been observed \cite{AVRON2} that the Kubo formula from which $\sigma_H$ is
obtained can be related to the integral of the first Chern class of some
Line Vector Bundle. This latter structure shows up because two parameters
have to be introduced within the quantum mechanical description of the considered systems.
Attempts to explain the observed fractional quantization of $\sigma_H$ have
also appeared, combining the above scheme with additional physical plausible
assumptions (e.g. the possible degeneracy of the ground state, the existence of a
gap above it and/or the validity of averaging $\sigma_H$ over the above two
parameters) [4-7].\par
The above "Topological Approach to the Quantum Hall Effect" has been
recognized as providing a possible explanation {\footnote{even when disorder
is included which however requires the actual validity of some additional
hypothesis}} for the observed robustness of the quantization of $\sigma_H$
but it is not quite fully satisfactory. It appears that this approach does
not permit one to really predict the value for $\sigma_H$ starting from
{\it{basic}} ingredients pertaining to the quantum mechanical description of
the systems. This may be due to the fact that the above mentionned Line
Vector Bundle does not capture (in most cases) a sufficient amount of the
physical properties shared by the real experiments (e.g. the ("engineering") 
geometry of the experimental devices). Notice that many quantum mechanical 
models used so far rely on assumptions made on the geometry of the system (e.g. starting
from a torus or annular geometry). While these assumptions are convenient
from a mathematical viewpoint (since they simplify the analyzis and/or permits
one to deal rather easily with a suitable choice for the boundary conditions), they are
not very natural from an experimental viewpoint. Ultimately, it would be
useful to incorporate into a single framework (hopefully) all the
geometrical and physical constraints characterizing the real experiments and
control the actual validity of the assumptions. This, together with a proper
inclusion of disorder, would permit one to describe in a single formalism 
integral and fractional quantization (and by the way to fully reconcile
localization with the topological machinery).\par
This talk summarizes some of the results derived in \cite{GMW1} and \cite{GMW2} in 
collaboration with Y. Georgelin and T. Masson. The physical system
we consider is a rectangle coupled to a four-point probe and is similar to
the one  considered in \cite{AVRON1}, \cite{NIU1}. It is depicted on figure 1. This is a natural choice
regarding the actual experimental devices as discussed in particular in
\cite{NIU1}, \cite{GMW1}. In most of the following discussion, the disorder is left
aside. 
In section 2, we outline the main
features of the formalism introduced in \cite{GMW1} which permits us to deal with
general boundary conditions for the physical system. Consistency of 
the necessary self-adjointness for the Hamiltonian description 
with general boundary conditions can only be achieved through the introduction
of two real parameters{\footnote{These parameters generalize the so called "boundary
condition angles" which appear e.g. when dealing with "Periodic or Toroidal
boundary conditions".}} (that will generate some "extended"
Brillouin zone). In section 3, we point out the occurence of a symmetry for the system which
can be identified with the irrational rotation algebra of the Non-Commutative
Torus \cite{RIEF}. This symmetry basically connects the physical space to the
"extended" Brillouin zone, that we call reciprocical space in the following. This symmetry 
is also briefly discussed in the light of the Non-Commutative Torus and
Area-Preserving Diffeomorphism algebra. In section 4, we review the case where
${{NB}\over{2\pi}}$ is integer. Here $N$ (resp. $B$) is the number of charge
carriers (resp. the external magnetic field) and ${{NB}\over{2\pi}}$ is 
the relevant control parameter for the system. The formalism of section 2 combined with the relevant
version of the irrational rotation algebra for ${{NB}\over{2\pi}}$ integer 
permits one to compute entirely $\sigma_H$ from the Kubo formula which 
is found to take integer or fractional values. In section 5, we discuss the case 
where ${{NB}\over{2\pi}}$ is not an integer for which the situation becomes far more 
complicated. When it takes rational values, the Hall conductivity can again 
be computed leading again to integer or fractional values. The extension of
the analyzis to the case where ${{NB}\over{2\pi}}$ is irrational does not lead to a reliable
characterization of the reciprocical space, therefore signaling the limitation of the
present approach. A possible way to circumvent this limitation (which could
also possibly lead to a proper incorporation of the disorder) is discussed
in section 6 where promising indications favoring the choice of a
more algebraic framework are breifly described.\par
\section{Hall system on a finite size sample}
We consider the system depicted on figure 1 where $N$ interacting spinless particles
with mass $m$ are confined on a rectangular sample (that we assume in the
following to be a square of unit length) submitted to an external
magnetic field $B$ and whose opposite edges are connected by wires. Units
are $e=\hbar=c=1$. 
$(x_i,y_i)$ (resp. $(p_{x_i},p_{y_i})$) denote the coordinates (resp. momenta) for 
particle $i$.\par
The electrons located in the planar sample are ruled by Quantum Mechanics
while the currents circulating in the wires (which are connected to
macroscopic devices in the experiments) should be regarded as classical
quantities. General boundary conditions can be obtained by constraining
these classical quantities. Namely, we assume that the currents circulating 
in the wires are conserved and
that the charge located on each edge of the rectangle is equal to the one
located on the corresponding opposite edge. These conditions will be
translated into constraints on the densities and currents expressed in the
Schr\"odinger representation. In this way, the presence of the wires
reflects itself as boundary conditions constraining the Quantum Mechanics
describing the dynamics of the electron in the sample.\par
\begin{figure}
\begin{center}
\includegraphics[scale=0.5] {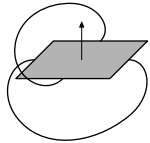}
\end{center}
\caption{The geometry of the system}
\end{figure}
As recalled in \cite{GMW1}, the Hamiltonian for the
system, in the absence of random potential, 
can be conveniently written in term of collective (center of mass) variables as
$$H={{1}\over{2Nm}}\big((P_x+{{1}\over{2}}NBy)^2+(P_y-{{1}\over{2}}NBx)^2\big)+H_I\equiv
H_0+H_I\eqno(1)$$
where $x={{1}\over{N}}\sum_{i=1}^Nx_i$, $P_x=\sum_{i=1}^Np_{x_i}$,
$y={{1}\over{N}}\sum_{i=1}^Ny_i$, 
$P_y=\sum_{i=1}^Np_{y_i}$ and $H_I$ depends only on the "internal" variables
(${\tilde{x}}_i=x_i-x$, ${\tilde{p}}_{x_i}=p_{x_i}-{{1}\over{N}}P_x$ and
similar expressions for the $y$-counterpart). In (1), the symmetric gauge
for the external gauge potential has been used. The corresponding Hilbert
space is then a tensor product ${\cal{H}}_0\otimes{\cal{H}}_I$ where the
collective (resp. internal) operators act only on ${\cal{H}}_0$ (resp.
${\cal{H}}_I$). In the following, we focus the discussion on the $H_0$ part
of the Hamiltonian together with the ${\cal{H}}_0$ part of the Hilbert space
since the observables we are interested in depends only on the center of mass variables 
and the relevant operators act only on ${\cal{H}}_0$.\par
To proceed further, one needs to built a representation of the abstract
operator algebra built from $H_0$, $P_x$, $P_y$. As we know from general
operator theory, care must be taken here because we work on a sample with finite 
size so that, in order to obtain a self-adjoint representation of the operators, one has to specify
carefully the space of functions on which they act. Therefore, requiring
self-adjointness and using boundary conditions constraints expressed as
constraints on the edge densities and currents in the Sch\"odinger
representation, we find the following result \cite{GMW1}:
$$P_x=-i\partial_x+{{NB}\over{2}}y\ ;\ P_y=-i\partial_y-{{NB}\over{2}}x
\eqno(2)$$
must act on the linear space of functions verifying
$$\Phi(1,y)=e^{i(\gamma+{{NB}\over{2}}y)}\Phi(0,y)\ ;\ 
\Phi(x,1)=e^{i(\eta-{{NB}\over{2}}x)}\Phi(x,0) \eqno(3)$$
where $\gamma$ and $\eta$ are two real arbitrary parameters which {\it{must}}
necessarely be introduced in the course of the construction and label
inequivalent representations for $P_x$ and $P_y$ \cite{GMW1} modulo $2\pi$. In
mathematical words, self-adjointness combined with boundary conditions preserving
current conservation in the wires and edge charge conservation force the
representation of the relevant operator algebra to be reducible. The space
generated by $\gamma$ and $\eta$ is called the {\it{reciprocical space}} 
and will be characterized more closely in a while. Note that, at this point,
$\gamma$ and $\eta$ are not constrained. Constraints on these parameters
will appear by further assuming specific values for ${{NB}\over{2\pi}}$.
According to the above discussion, the Hilbert space
${\cal{H}}_0$ is a direct sum (actually a direct Hilbertian integral) of
each Hilbert space indexed by each pair $(\gamma,\eta)$ and from (3) any
general wave function $\psi(x,y;\gamma,\eta)$ must satisfy
$$\psi(1,y;\gamma,\eta)=e^{i(\gamma+{{NB}\over{2}}y)}\psi(0,y;\gamma,\eta)\ ;
\ \psi(x,1;\gamma,\eta)=e^{i(\eta-{{NB}\over{2}}x)}\psi(x,0;\gamma,\eta) \eqno(4).$$
The various differences between the present framework and other related
works are discussed in the section 4 of \cite{GMW1} to which we refer for
more details. Before going further, some comments are in order. Equation (4)
defines $\psi$ as being a section of a Line Vector Bundle over a Torus in
the $(x,y)$-space {\it{only when}} ${{NB}\over{2\pi}}$ is integer. In this case, the
corresponding base space, a $(x,y)$-Torus with periodicity 1, has nothing to
do with the physical sample (whose topology is that of a Torus with one
puncture). When ${{NB}\over{2\pi}}$ takes rational values, the base space
(in the $(x,y)$-space) of
the Line Vector Bundle becomes a Rieman surface of higher genus with
punctures \cite{GMW2}. 
\section{The irrational rotation algebra}
It appears that the system built from $H_0$ and ${\cal{H}}_0$ has a
symmetry \cite{GMW1}, \cite{GMW2}. It is generated by a family of unitary operators
${\cal{U}}_{\sigma,\theta}$ indexed by two parameters $\sigma$ and $\theta$
verifying
$${\cal{U}}_{\sigma,\theta}\psi(x,y;\gamma,\eta)=
e^{-{{i}\over{2}}NB(\sigma
x+\theta y)}\psi(x+\theta,y-\sigma;\gamma+NB\sigma,\eta+NB\theta) \eqno(5)$$
$${\cal{U}}_{\sigma_1,\theta_1}{\cal{U}}_{\sigma_2,\theta_2}=
e^{-iNB(\sigma_2\theta_1-\sigma_1\theta_2)}
{\cal{U}}_{\sigma_2,\theta_2}{\cal{U}}_{\sigma_1,\theta_1} \eqno(6)$$
where (5) holds provided the wave functions admit
(monovalued) extensions on some suitable space related to $R^2$, a condition which can be
always achieved \cite{GMW2}. Here, we point out that this symmetry connects
both the $(x,y)$ space and the reciprocical space as it can be
realized from (5). It appears that this observation proves usefull in the
computation of the Hall conductivity when ${{NB}\over{2\pi}}$ takes
integer values. Equation (6) is reminiscent of the magnetic algebra obeyed by 
the magnetic translation operators in Landau type problems. In fact, when 
${{NB}\over{2\pi}}$ is irrational, (6) is known in the
mathematical litterature as the "irrational rotation algebra" of the
Non-Commutative Torus \cite{RIEF}. It appears that it can be related to the
abstract algebra defining the Non-Commutative Torus. This abstract algebra
is defined as an associative (non commutative) unital $*$-algebra with
involution $\dag$ generated by two elements $Z_1$ and $Z_2$ satisfying
$$Z_1Z_2=e^{i2\pi\Theta}Z_2Z_1\ ,\ Z_i^\dag=Z_i^{-1},\ i=1,2\ ,\
\Theta\in[0,1] \eqno(7),$$
($\Theta$ is usually called the non commutativity parameter) with
smooth completion obtained from all elements $f$ of the form{\footnote{which
in some sense generalizes the usual Fourier series expansion}}
$$f=\sum_{m_1,m_2\in Z^2}f_{(m_1,m_2)}e^{i\pi\Theta
m_1m_2}Z_1^{m_1}Z_2^{m_2}  \eqno(8)$$
where the (Schwartz sequences) $f_{(m_i,m_j)}$'s decrease sufficiently fast for
$|m_i|\to\infty$. One observes that (6) reduces to (7) upon setting
$Z_1\equiv{\cal{U}}_{\sigma,0}$, $Z_2\equiv{\cal{U}}_{0,\theta}$ and
$NB\sigma\theta=2\pi\Theta$. Notice that, by defining $T_{(n_1,n_2)}$
$\equiv{{1}\over{\Theta}}e^{in_1n_2\Theta/2}Z_1^{n_1}Z_2^{n_2}$, one obtains
from (8)
$$[T_{(n_1,n_2)},T_{(m_1,m_2)}]={{i2}\over{\Theta}}\sin(\Theta(n_1m_2-n_2m_1))
T_{(n_1+m_1,n_2+m_2)} \eqno(9).$$
This infinite dimensional Lie algebra is the "trigonometric algebra"
initially advocated in \cite{FAIR} which, in the limit $\Theta\to0$ reduces to
$[T_{(n_1,n_2)},T_{(m_1,m_2)}]=$$i2($\hfill\break
$n_1m_2-n_2m_1)T_{(n_1+m_1,n_2+m_2)}$,
that is the $W_\infty$ algebra of the Area-Preserving Diffeomorphisms of
the Torus.\par
\section{Computing the Hall conductivity}
When ${{NB}\over{2\pi}}=l$, $l$ integer, that we call the integer case, 
we find that (4) can be extended as
$$\psi(x+1,y;\gamma,\eta)=e^{i(\gamma+{{NB}\over{2}}y)}\psi(x,y;\gamma,\eta)
\eqno(10a),$$
$$\psi(x,y+1;\gamma,\eta)=e^{i(\eta-{{NB}\over{2}}x)}\psi(x,y;\gamma,\eta)
\eqno(10b)$$
which, for fixed $(\gamma,\eta)$, permits us to interpret $\psi$ as a
section of a Line Vector Bundle (LVB) over a torus in the $(x,y)$-space with
periodicity 1. Notice that the quantization condition ${{NB}\over{2\pi}}=l$
simply expresses the fact that the first Chern class of this LVB, ${{NB}\over{2\pi}}$ 
, is an integer $l$. Recall that the above torus has nothing
to do with the physical sample! It is convenient at this point to perform the following
gauge transformation ${\tilde{\psi}}(x,y;\gamma,\eta)$$=e^{-i(\gamma x +\eta
y)}\psi(x,y;\gamma,\eta)$. The relevant symmetry stemming from (5) and
(6) is generated by 
$${\cal{U}}_{1/l,0}\equiv {\tilde{U}}\ ,\ {\cal{U}}_{0,1/l}\equiv{\tilde{V}}
\ ;\ {\tilde{V}}{\tilde{U}}=e^{i{{2\pi}\over{l}}}{\tilde{U}}{\tilde{V}}
\eqno(11)$$
and the (gauge transformed) wave functions ${\tilde{\psi}}$
can be splitted into $l$ components ${\tilde{\psi}}_I$, $I=1,...,l$ which,
upon using (10), (11) and (5) in which the ${\cal{U}}$'s are replaced by their
${\tilde{U}},{\tilde{V}}$ counterparts, must satisfy
($q=e^{i{{2\pi}\over{l}}}$)
$${\tilde{\psi}}_I(x,y;\gamma,\eta)=q^{1-I}e^{i(\pi
y+{{\gamma}\over{l}})}{\tilde{\psi}}_I(x+{{1}\over{l}},y;\gamma,\eta+2\pi)
\eqno(12a)$$
$${\tilde{\psi}}_{I+1}(x,y;\gamma,\eta)=e^{i(\pi
x+{{\eta}\over{l}})}{\tilde{\psi}}_I(x,y-{{1}\over{l}};\gamma+2\pi,\eta)\eqno(12b)$$
and
$${\tilde{\psi}}_I(x,y;\gamma+2\pi l,\eta)=e^{-i2\pi
lx}e^{i\eta}{\tilde{\psi}}_I(x,y;\gamma,\eta) \eqno(13a),$$ 
$${\tilde{\psi}}_I(x,y;\gamma,\eta+2\pi l)=e^{-i2\pi
ly}e^{-i\gamma}{\tilde{\psi}}_I(x,y;\gamma,\eta) \eqno(13b).$$
The relation (12) expresses the fact that the function ${\tilde{\psi}}_I$
(resp. ${\tilde{\psi}}_{I+1}$) restricted to the domain $\eta\in[2\pi
n,2\pi(n+1)]$, $1\le n\le l-1$ (resp. $\gamma\in[2\pi m,2\pi(m+1)]$, $1\le
m\le l-1$) is completely determined by the restriction of ${\tilde{\psi}}_I$ on $\eta\in[0,2\pi]$
(resp. $\gamma\in[2\pi(m-1),2\pi m]$. More importantly, the relation (13),
obtained from (10) with the help of (11), permits us  
to interpret each ${\tilde{\psi}}_I$ as a section of a LVB over a
torus in the $(\gamma,\eta)$ reciprocical space with periodicity $2\pi l$.\par
Now we can use the Kubo formula to compute the Hall conductivity $\sigma_H$. By using
standard manipulations, $\sigma_H$ can be expressed as \cite{GMW1} 
$$\sigma_H={{N^2}\over{i(2\pi)^2}}{{1}\over{l}}\sum_{I=1}^l\int_{[0,2\pi]^2}d\gamma
 d\eta\int_{[0,1]^2}dxdy\big({{\partial{\tilde{\psi}}^*_I}\over{\partial\gamma}}
{{\partial{\tilde{\psi}}_I}\over{\partial\eta}}-{{\partial{\tilde{\psi}}^*_I}\over{\partial\eta}}
{{\partial{\tilde{\psi}}^*_I}\over{\partial\gamma}}\big) \eqno(14).$$
Set now
$$\Omega_I\equiv\int_{[0,1]^2}dxdy\big({{\partial{\tilde{\psi}}^*_I}\over{\partial\gamma}}
{{\partial{\tilde{\psi}}_I}\over{\partial\eta}}-{{\partial{\tilde{\psi}}^*_I}\over{\partial\eta}}
{{\partial{\tilde{\psi}}^*_I}\over{\partial\gamma}}\big),\ I=1,...,l\eqno(15).$$
By further making use of (12), one can show \cite{GMW1} that
$${{1}\over{l}}\sum_{I=1}^l\int_{[0,2\pi]^2}d\gamma
 d\eta\Omega_I={{1}\over{l^2}}\int_{[0,2\pi l]^2}d\gamma
 d\eta\Omega_1 \eqno(16)$$
where in the RHS of (16), one can easily recognizes \cite{GMW1} the first Chern number
for the relevant LVB over the Torus in the reciprocical space with periodicity $2\pi l$,
namely $C_1^R=i2\pi\int_{[0,2\pi l]^2}d\gamma d\eta\Omega_1$. It appears
that this Chern number can be computed entirely. The derivation can be found
in the appendix B of \cite{GMW1} and the result is $C_1^R=i4\pi l$. This, combined
with (16) yields
$$\sigma_H={{1}\over{2\pi}}({{2N^2}\over{l}}) \eqno(17).$$
Therefore, when ${{NB}\over{2\pi}}=l$, $l$ integer, we find that the Hall
conductivity can take integer or fractional values.\par
\section{The non-integer case}
The situation becomes far more complicated when ${{NB}\over{2\pi}}$ is not
an integer value. When ${{NB}\over{2\pi}}=l/k$, the Hall conductivity
$\sigma_H$ can again be explicitely computed \cite{GMW2}, provided some additional
assumptions on the initial interaction potential is made. We find that
$\sigma_H$ takes again integer or fractional values, namely
$$\sigma_H={{1}\over{2\pi}}{{2(kN)^2}\over{kl}} \eqno(18).$$
If the above assumption is relaxed, one has first to extend (4) to some suitable
domain of the $(x,y)$-space built from $R^2$. We find that this can be done on a Riemann surface
${\cal{S}}_k$ with genus $g={{1}\over{2}}(2+k^2(k-1))$ with $k^2$ punctures. The proof (given
in appendix A of \cite{GMW2}) is rather involved and amounts to combine surface surgery to the van Kampen
theorem. Roughly speaking, ${\cal{S}}_k$ is obtained by gluing $k^3$ copies of the 
physical sample and receives a magnetic flux equal to $2\pi lk^2$, each puncture carrying a
magnetic flux equal to $2\pi l$. The relation (4) (with the gauge transformed wave functions 
${\tilde{\psi}}$ as defined in section 4) becomes 
$${\tilde{\psi}}(x+k,y;\gamma,\eta)=e^{ik\gamma+{{i}\over{2}}2\pi ly}
{\tilde{\psi}}(x,y;\gamma,\eta) \eqno(19a),$$
$${\tilde{\psi}}(x,y+k;\gamma,\eta)=
e^{ik\eta-{{i}\over{2}}2\pi lx}{\tilde{\psi}}(x,y;\gamma,\eta) \eqno(19b),$$
so that it can be now interpreted a section of a LVB over ${\cal{S}}_k$ in
the $(x,y)$-space.
The rest of the computation of $\sigma_H$, although very involved, could be done in principle
by relating the above LVB to its counterpart in the reciprocical space
through (5) and (6) and using the representations of the relevant algebra
stemming from (6). When 
${{NB}\over{2\pi}}$ takes irrational values, the relevant space on which (4)
can be extended is very complicated: it is the universal covering space of
$R^2/Z^2$ \cite{GMW2} and no conclusion can be obtained from the present formalism.\par
\section{Discussion and outlook}
Provided some reliable identification of the reciprocical space (i.e the
Brillouin zone) can be done, integer as well as fractional quantization for
$\sigma_H$ can be shown within the present approach. This latter however
loses its efficiency when no Brillouin zone can be characterized which is
the case when ${{NB}\over{2\pi}}$ is irrational or when the disorder is
taken into account. We note that plausible
phenomenological conclusions can be obtained in the presence of (at least
weak) disorder by supplementing the present
analyzis with reasonable physical assumptions somehow similar to those 
initialy proposed in \cite{WU}. We will not discuss this
point here, refering to \cite{GMW2} for more details. The conclusions we obtain are
consistent either with the physics stemming from the global phase diagram
proposed in \cite{KLZ} or with the phase diagram that we proposed in
\cite{PLA}.\par
The Chern numbers are not the only topological invariants playing a salient
role in the Topological approach for the Quantum Hall Effect. When no
Brillouin zone can be identified, it appears
that $\sigma_H$ can also be related to another topological invariant
\cite{BELSS}, namely the Fredholm index which is stable under (small) deformations and can be
viewed as some non-commutative version of the Chern number. The introduction
of Fredholm index permits one to deal with disorder but only integer
quantization for $\sigma_H$ is recovered. So far, a self-contained
mathematical framework dealing with interacting electrons in the presence of
a random potential (in a realistic geometry similar to the one depicted on
fig.1) and giving rise to integer as well as fractional quantization for
$\sigma_H$ is still missing.\par
The analyzis presented in \cite{GMW1}, \cite{GMW2} involves encouraging
indications favoring a more algebraic approach that could reach this goal
and appears to be related to the Kasparov K-K Theory \cite{SKAN}. To get a
flavor of this,  one first notices that deformations can be better treated
within an algebraic framework. One further notices that LVB can be indeed
completely characterized by their sections. This stemms basically from the
Serre-Swan Theorem arising in Bundle Theory which asserts that there is a
one-to-one correspondance between LVB over some base space and projective
modules over the algebra of functions describing this base space. Projective
modules are classified by K-Theory. Now, sections in the $(x,y)$-space must
be connected to their counterpart in the reciprocical space through
transformations associated with the irrational rotation algebra. This
amounts to connect K-theory in the $(x,y)$-space to its K-Theory counterpart
in the reciprocical space through the above transformations which is
reminiscent of (Kasparov) K-K Theory. The actual implementation of
these ideas in Quantum Hall systems as well as in other physical situations
are presently under study \cite{TW}.\par
{\bf{Acknowledgements}}: The content of this talk has been greatly
influenced by many stimulating discussions with my collaborators Y.
Georgelin and T. Masson. I am grateful to the organizers of the "Conference on Higher Dimensional
Quantum Hall Effect, Chern-Simons Theory and Non-commutative Geometry in
Condensed Matter Physics and Field Theory" for invitation. Hospitality of
ICTP (Trieste) is also acknowledged.\par

\end{document}